\documentclass[reprint, amsmath,amssymb, aps]{revtex4-2}

\usepackage{graphicx}
\usepackage{subfigure}
\usepackage{multirow}
\usepackage{xcolor, soul}

\usepackage{longtable}
\usepackage[T1]{fontenc}
\usepackage{dcolumn}

\usepackage{bm}
\usepackage{amsfonts}
\usepackage{amsmath}
\usepackage{amssymb}
\usepackage{blindtext}
\usepackage{textgreek}

\begin{document}

\title{Analysis of optical loss thresholds in the fusion-based quantum computing architecture}

\author{Aleksandr Melkozerov$^{1}$}
\email{melkozerov.aa17@physics.msu.ru}
\author{Ashot Avanesov$^{2}$}
\author{Ivan Dyakonov$^{1,2}$}
\author{Stanislav Straupe$^{1,2}$}
\affiliation {\it $^1$Quantum Technology Centre and Faculty of Physics, M.\,V. Lomonosov Moscow State University, 1 Leninskie Gory, Moscow, 119991, Russia \\
\it $^2$Russian Quantum Center, 30 Bolshoy Boulevard, building 1, Moscow, 121205, Russia
}

\date{April 2, 2024}

\begin{abstract} 
Bell state measurements (BSM) play a significant role in quantum information and quantum computing, in particular, in fusion-based quantum computing (FBQC). The FBQC model is a framework for universal quantum computing provided that we are able to perform entangling measurements, called fusions, on qubits within small entangled resource states. Here we analyse the usage of different linear-optical BSM circuits as fusions in the FBQC schemes and numerically evaluate hardware requirements for fault-tolerance in this framework.
We examine and compare the performance of several BSM circuits with varying additional resources and estimate the requirements on losses for every component of the linear-optical realization of fusions under which errors in fusion networks caused by these losses can be corrected. Our results show that fault-tolerant quantum computing in the FBQC model is possible with currently achievable levels of optical losses in an integrated photonic implementation, provided that we can create and detect single photons of the resource states with a total marginal efficiency higher than 0.973.

\end{abstract}

\maketitle 

\section{Introduction}

A Bell state measurement is an operation of paramount importance in many linear-optical quantum protocols. It finds applications in quantum communication \cite{Duan2001}, including quantum teleportation and entanglement swapping \cite{Bennett1993, Zukowski1993}, and in quantum computing \cite{Browne2005, Raussendorf2003, Raussendorf2006, Kieling2007, GimenoSegovia2015}. It also acts as a central part of fusion-based quantum computing \cite{Bartolucci2023}, recently proposed architecture for fault-tolerant quantum computing.

Non-classical properties of single photons interfering in linear-optical elements lay the foundation of a linear-optical platform, which is considered as a perspective candidate for the implementation of a quantum computer \cite{Knill2001, OBrien2007, Carolan2015, Kok2010}. However, optical losses put serious constraints on practical schemes since in principle they cannot be completely eliminated \cite{Li2015, Ralph2005, Lee2019}. Moreover, deterministic entangling operations with photonic qubits in dual-rail encoding are prohibited in a linear optical system \cite{Calsamiglia2001}, and probabilistic nature of computation generally results in a large resource and technical overhead.
These problems can be solved by error correction procedures, available in computation models different from standard circuit model \cite{Raussendorf2006, Bartolucci2023}, providing the path to scaling up optical quantum computing.

In the FBQC model quantum algorithms are decomposed into a sequence of two repeated fundamental operations:

\begin{enumerate}
    \item Generation of constant-sized entangled states called \textit{resource states}.
    \item Projective entangling measurements, called \textit{fusions}, such as Bell state measurements.
\end{enumerate}

The quality and success probabilities of both these operations are extremely important for the successful realization of FBQC. In this work, we study the effect of losses on the fusion operations and estimate the limits required to achieve large-scale fault-tolerant computations.

It is known that the success probability of a BSM in the dual-rail encoding cannot exceed 50\% using linear-optical elements, arbitrarily many ancillary vacuum modes, and photon number resolving detectors (PNRDs) \cite{Michler1996, Ltkenhaus1999}. However, the success probability can be elevated at the cost of extra photons in more sophisticated linear-optical schemes. 

In particular, using ideas from \cite{Grice2011} and \cite{Ewert2014}, we investigate linear-optical BSM circuits yielding higher success probabilities with different additional resources. We numerically estimate the requirements on the performance of linear-optical constituents of various BSM circuits used as fusions in the FBQC model. Specifically, we examine two notable fusion networks introduced in the seminal work~\cite{Bartolucci2023}, as well as some modification of these networks, and derive threshold values for losses from different sources in the BSM circuits, under which errors caused by these losses are tolerated. In addition, we compare the performance of the circuits with different additional resources and find a trade-off between the improvement from the success probability boosting with additional photons and levels of losses imposed by additional linear-optical components in the more complicated BSM circuits.

The paper is organized as follows. Sec. II details some preliminaries of the work. Sec. III provides a detailed explanation of the properties of linear-optical realization of Bell state measurements and some possible explicit 
schemes of BSM circuits with different additional resources. In Sec. IV we give a brief recap of the FBQC model and an overview of the fusion networks to be considered. Sec. V discusses sources of imperfections in linear optics and the model of losses used in this work. In Sec. VI we show the results of numerical simulations of the lossy BSM circuits within the FBQC framework including threshold hardware requirements for the components of linear-optical circuits for fault-tolerant quantum computation. Finally, in Sec. VII we give an overview of the results and the conclusion.

\section{Dual-rail encoding}
Throughout this work we consider a dual-rail encoding of photonic qubits, which is the most convenient and common encoding in optical quantum computing. A logical qubit here is represented by a photon occupying either of two modes: $|0\rangle_L = |10\rangle = h^{\dagger}|vac\rangle$ and $|1\rangle_L = |01\rangle=v^{\dagger}|vac\rangle$, where $h^{\dagger}$ and $v^{\dagger}$ are the creation operators acting on the first and the second mode respectively, $|vac\rangle = |0\rangle^{\otimes N}$.
We focus on path encoding, where the two modes used to encode the computational basis states of a single qubit are two different spatial channels. This is essentially equivalent to the common polarization qubit encoding, where the two modes are the internal polarization degrees of freedom of the photon, and these two representations can be deterministically converted into each other \cite{https://doi.org/10.25560/43936}.

\begin{figure}[h]
    \includegraphics[width=3.3in]{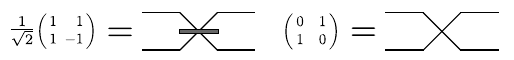}
    \caption{Ideal lossless building blocks. A line corresponds to an optical mode. A two-channel $50:50$ beamsplitter is described by a $2\times2$ Hadamard matrix.}
    \label{fig:basic_elements}
\end{figure}

Arbitrary lossless $N$-channel linear-optical interferometer performs a photonic state transformation which may be described by an $N \times N$ unitary matrix $U$ acting on creation or annihilation operators of the input modes. In this paper we present linear-optical schemes composed of two simple elements only: a two-channel $50:50$ beamsplitter and a mode swap. These elements can be easily implemented, for example, in integrated photonics \cite{Chrostowski2015, Johnson2020}. Fig. \ref{fig:basic_elements} introduces their depiction and corresponding transfer matrices.

\section{Bell state measurement}
A BSM is a projection of two qubits onto one of the four Bell states. BSM can also be described as a joint measurement of two Pauli operators $X_1 X_2$ and $Z_1 Z_2$ on the state of two input qubits, where $X_i$ and $Z_i$ are single-qubit Pauli $X$ and $Z$ operators acting on $i$-th qubit. Projection into different Bell states corresponds to the outcomes of the Pauli operators' measurement as follows:

\begin{equation*}
    \begin{aligned}
        & \left|\Phi^{ \pm}\right\rangle_L=\frac{1}{\sqrt{2}}\left(\left|0_1 0_2\right\rangle \pm\left|1_1 1_2\right\rangle\right)_L \Leftrightarrow\left\{\begin{array}{l}
            Z_1 Z_2 \rightarrow+1 \\
            X_1 X_2 \rightarrow \pm 1
        \end{array}\right. ,\\
        & \left|\Psi^{ \pm}\right\rangle_L=\frac{1}{\sqrt{2}}\left(\left|0_1 1_2\right\rangle \pm\left|1_1 0_2\right\rangle\right)_L \Leftrightarrow\left\{\begin{array}{l}
            Z_1 Z_2 \rightarrow-1 \\
            X_1 X_2 \rightarrow \pm 1
        \end{array}\right. .
    \end{aligned}
\end{equation*}

In the following the subscripts denoting the qubit on which the Pauli operator is acting on will be omitted for simplicity unless it causes confusion.

\begin{figure}[t]
\includegraphics[width=3.3in]{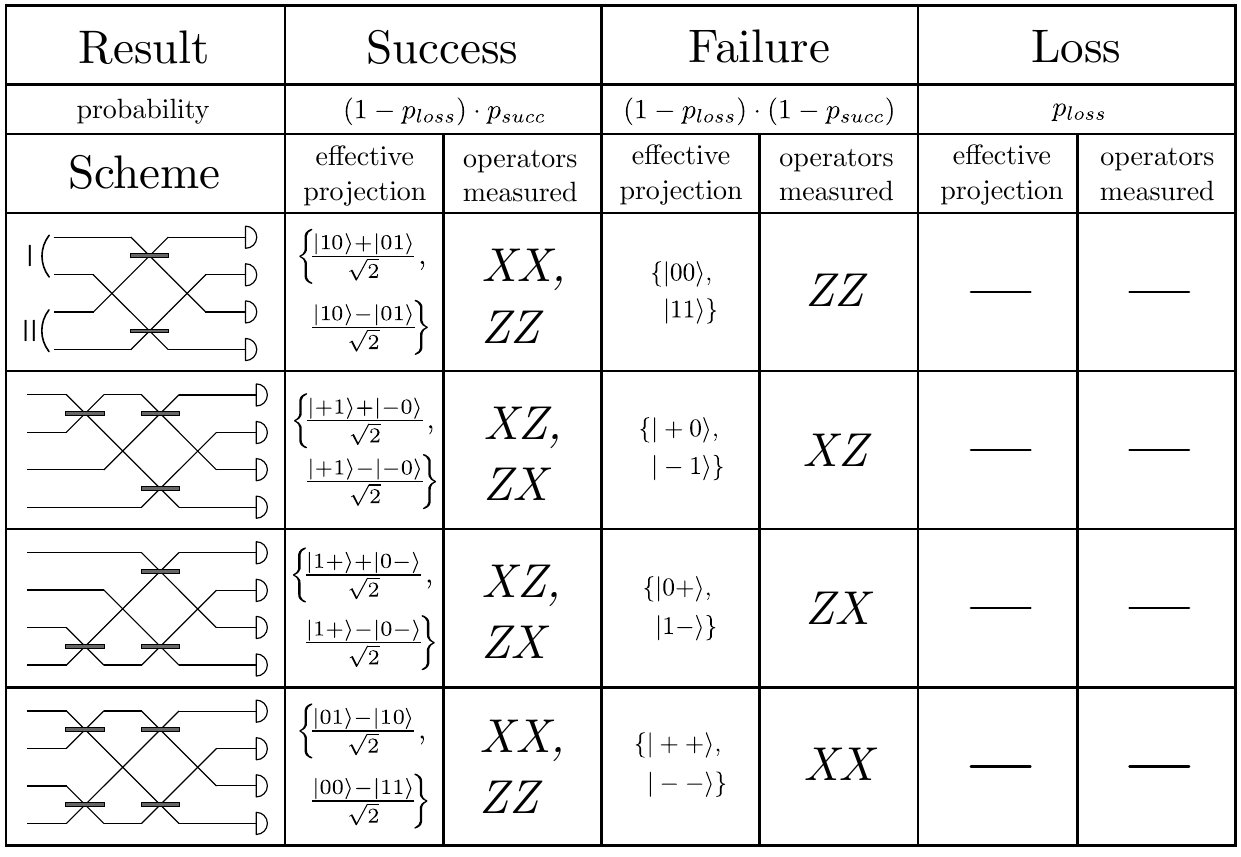}
\caption{\label{fig:table} Possible linear-optical circuits for the Bell state measurement. The first of the two initial dual-rail encoded qubits occupy the upper two modes, second -- the bottom two modes as denoted by I and II in the first picture respectively. Each scheme projects initial qubits in the subspace of states depicted in the logical state notation. All circuits here have the same success probability of $p_{succ} = 0.5$.}
\end{figure}

\subsection{Linear-optical BSM}\label{lobs}

In linear-optical quantum computing a variant of probabilistic linear-optical BSM is known as a Type-II fusion gate, which was initially introduced in \cite{Browne2005}.
In general, successful Type-II fusion "connects" two entangled states of the form 
\begin{equation}
\frac{1}{\sqrt{2}}\left(\left|A_0\right\rangle|0\rangle+\left|A_1\right\rangle|1\rangle\right)_L \otimes \frac{1}{\sqrt{2}}\left(\left|B_0\right\rangle|0\rangle+\left|B_1\right\rangle|1\rangle\right)_L,
\end{equation}
leaving the system in the state $\left|A_0 B_0\right\rangle_L \pm\left|A_1 B_1\right\rangle_L$ or $\left|A_1 B_0\right\rangle_L \pm\left|A_0 B_1\right\rangle_L$ at the cost of destroying the two measured qubits. This operation allows for loss tracking if realized with a linear-optical circuit in which all four modes of the two input qubits are measured by photon number resolving detectors (PNRDs). If PNRDs count less than two photons in total in four modes of two initial qubits, we can say that photon loss occurred at some stage of the scheme.

Even in lossless linear-optical schemes an unambiguous BSM cannot be realized deterministically \cite{Vaidman1999, Calsamiglia2001} assuming that the two initial qubits are from separate states. So every linear-optical realization of a BSM has a finite success probability in the lossless case $p_{succ}$.

Type-II fusions may differ by the effective projection they preform, and, therefore, by the Pauli operators they measure \cite{https://doi.org/10.48550/arxiv.2312.02377}. Some Type-II fusion linear-optical circuits based on BSM and their properties are depicted in the Fig.~\ref{fig:table}. Each scheme "succeeds" when no more than one photon is measured at each detector and "fails" when both photons are detected in the same mode.
In the success case the input qubits are projected to one of the Bell states and intended operators $XX$ and $ZZ$ are measured. Rearranging the scheme, we can change the basis of the measurement and effective projection.

\begin{figure*}[ht]
\includegraphics[width=7in]{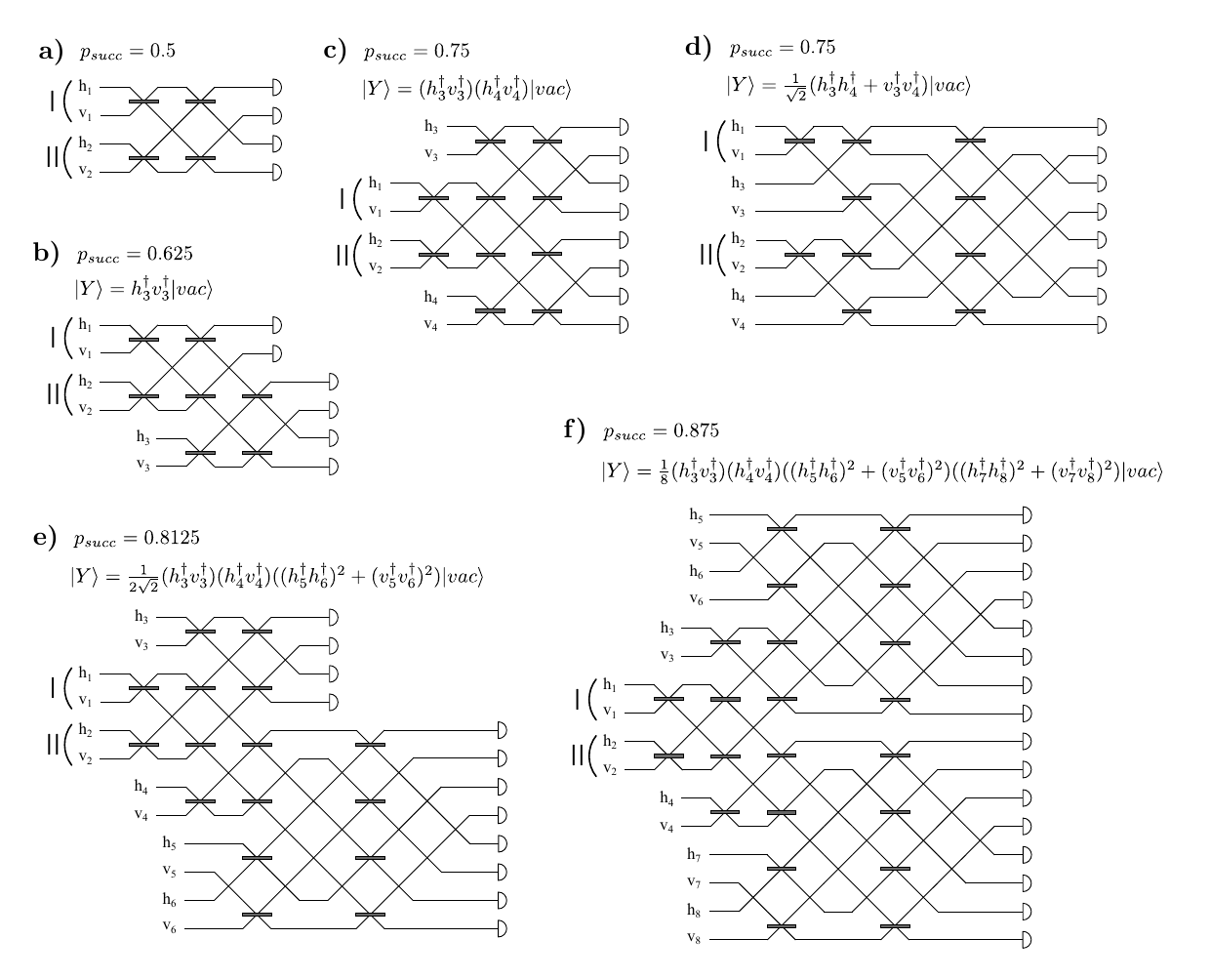}
\caption{\label{bsm_table} a) Regular BSM circuit. b)-f) possible boosted versions of the BSM circuit a). Schemes differ by the success probability in the lossless case $p_{succ}$ and corresponding ancillary states $|Y\rangle$ used for boosting. In each case two initial dual-rail encoded qubits that we want to project onto Bell states occupy modes $h_1$, $v_1$ and $h_2$, $v_2$, which are denoted as I and II in the picture respectively. All schemes measure operators $XX$ and $ZZ$ in the case of success and the $XX$ operator in the case of failure.}
\end{figure*}

The fusion fails with probability $p_{fail} = 1-p_{succ}$. The failure event is equivalent to separable single-qubit measurements. Depending on the circuit used to perform the fusion, the measured operators can be $XI$ and $IX$, or $ZI$ and $IZ$. Receiving the results of $XI$ and $IX$ measurements, we can reconstruct the outcome of one of the initially desired two-qubit measurements, $XX$, but not $ZZ$, and vice versa. Fault-tolerant frameworks such as FBQC make use of this property so that the failure outcomes are taken into account by error correction codes.

The exact mapping between the detection results and effective projections can be understood by the evolution of the detected states in the reversed scheme to see which initial states have generated them \cite{https://doi.org/10.25560/43936}.  

If any photon is lost in the scheme, we do not obtain any results of the Pauli measurements at all. Let the probability of any loss in a scheme be $p_{loss}$. Then we can identify three possible outcomes of a linear-optical fusion: 
\begin{enumerate}
  \item \textit{Success.} The $XX$ and $ZZ$ operators are measured, occurs with probability $(1-p_{loss}) p_{succ}$. 
  \item \textit{Failure.} The $XX$ or $ZZ$ operator is measured, occurs with probability $(1-p_{loss}) (1-p_{succ})$. 
  \item \textit{Loss.} No operators measured, occurs with probability $p_{loss}$. 
\end{enumerate}

Circuits in Fig.~\ref{fig:table} reach the success probability limit of $p_{succ} = 0.5$ for the BSM using passive linear-optical elements, arbitrarily many vacuum ancilla modes, PNRDs \cite{Michler1996, Calsamiglia2001}, and no additional ancillary photons. However, one can use additional photonic states to increase $p_{succ}$.

\subsection{Boosted BSM} \label{Boosted BSM}

The so-called boosted schemes use extra ancillary photons at the input to upscale the linear-optical BSM success probability, see Fig.~\ref{bsm_table} (b-f). Two main boosting techniques were presented in \cite{Grice2011} and \cite{Ewert2014}. The first approach deals with additional two-photon Bell state to boost the probability of successful BSM to $p_{succ} = 0.75$, while the second approach achieves the same improvement by using four initially unentangled photons.

These methods can be generalized by adding more ancillary photons to reach higher success probabilities. For example, in the first approach additional $(2^{k} - 2)$ entangled ancillary photons yield a success probability of $p_{succ} = 1 - (1/2)^k$. However, this improvement requires entangled states of a larger number of photons in both methods.

In Fig.~\ref{bsm_table} we present possible boosted versions (b-f) of the linear-optical BSM circuit (a). We use both known techniques to boost the probability of success $p_{succ}$, along with the modification of the second technique, that allows us to reach intermediate success probabilities, and heuristically optimize the total number of beamsplitters in each circuit. Schemes differ by $p_{succ}$ and ancillary states $|Y\rangle$ used for their boosting. Two initial qubits in dual-rail encoding that we want to project onto the Bell basis occupy modes $h_1$, $v_1$ and $h_2$, $v_2$ respectively.
Ancillary states $|Y\rangle$ are launched in other modes, and their states are presented in the mode creation operator's notation in the Fig.~\ref{bsm_table},
where $h^{\dagger}_i$ corresponds to the creation of a photon in the $h_i$ circuit mode, same for $v^{\dagger}_i$. Each circuit measures $XX$ and $ZZ$ operators in the case of success, and $XX$ in the case of failure. They can be modified to measure other operators by removing up to two beamsplitters acting on modes $h_1$, $v_1$ and $h_2$, $v_2$ at the beginning of the scheme, similarly to regular BSMs (see Fig. \ref{fig:table}).

For practical applications the preferable circuits are those which contain the least number of trivial elements.
For convenience of analysis the circuits shown in  Fig.~\ref{bsm_table} include mode swaps so the initial qubits are put in the interferometers one by one, second mode after the first, and detectors measure the corresponding initial modes. In practice, most of these mode swaps may be removed by simple rearrangement of the initial modes and the detectors at the output of each circuit, and they are not accounted as the sources of losses in further investigations.

\section{Recap: FBQC}\label{sec:FBQC}

\begin{figure}[t]
\includegraphics[width=3.2in]{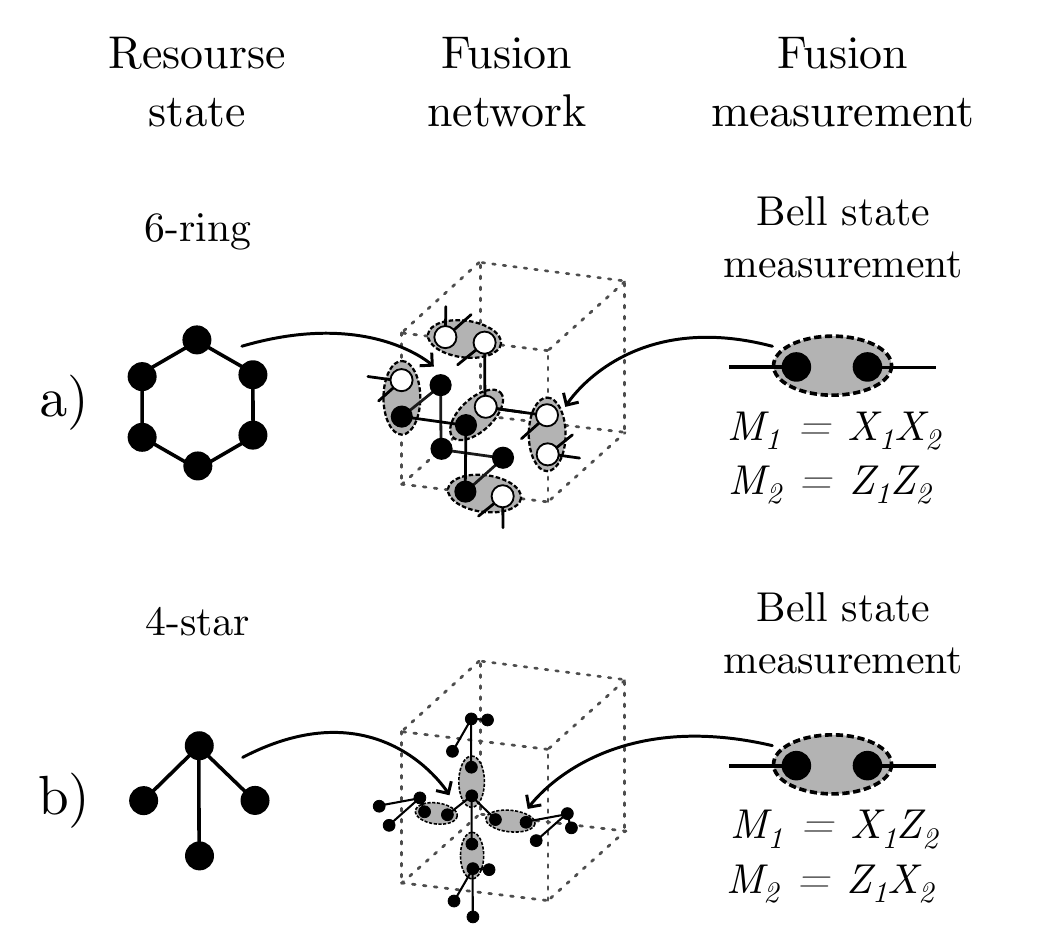}
\caption{\label{fig:fusion_networks} a) 6-ring fusion network, b) 4-star fusion network from \cite{Bartolucci2023}. Resource states are presented in the graph representation up to single-qubit operations.}
\end{figure}

We study the performance of different BSM schemes within a fusion-based quantum computing model \cite{Bartolucci2023}. The original paper develops a theory of fault-tolerant quantum computing based on small entangled resource states and entangling fusion measurements. This framework is highly suitable for linear-optical implementation because it supposes short qubit lifetime and can correct errors caused by the non-determinism of fusions and optical losses as long as this errors occur with sufficiently low probability.

The computation in the FBQC model resides on building a network of resource states and fusion measurements between them which is called a fusion network. A fusion network defines the amount of resources for computation and the loss tolerance. Here we will consider networks, shown in Fig.~\ref{fig:fusion_networks}, which were proposed in \cite{Bartolucci2023}.

Both fusions and resource states in fusion networks can be described using a stabilizer formalism. 
The key for fault-tolerance in FBQC is redundancy between measurement stabilizers and resource state stabilizers, which is captured by the check operator group. The check operator group generates a syndrome graph, where edges represent single fusion outcomes and vertices represent parity checks.

In an ideal situation every fusion measurement produces two measurement outcomes. In general, all errors in a fusion network may be described by an erasure of each individual outcome or it's flip regardless of the hardware implementation.
In terms of linear optical BSM, we can say that when BSM circuit acts successfully, both measurement outcomes are obtained, in the case of failure one outcome is erased, and both outcomes are erased in the case of loss, see Section~\ref{lobs}.

Let the combined probability of a single measurement outcome erasure due to any imperfection be $p_{er}$. If we suppose that every measurement outcome is erased with probability $p_{er}$,  there exists a marginal threshold erasure probability $p_{er}$, below which the fusion network tolerates losses-induced errors. For the networks in Fig.~\ref{fig:fusion_networks}, these threshold $p_{er}$ are 6.90\% for the 4-star network and 11.98\% for the 6-ring network \cite{Bartolucci2023}. Note that these probabilities are marginal and calculated in the absence of outcome flips.

For a linear-optical realization of the networks we will consider a static bias arrangement. This means that circuits for fusion are chosen randomly so that with the 0.5 probability fusion measures only the $ZZ$ operator in the case of failure, and only the $XX$ operator in another failure case with the same probability. We denote the probability of a single measurement outcome erasure $p_{er}$ in this specific linear-optical setup as $p_0$. Then, one of the measurement outcomes $XX$ or $ZZ$ is erased in the case of loss which occurs with probability $p_{loss}$, and with probability $(1-p_{loss})(1-p_{succ})/2$ in the case of failure.
Therefore, for a single outcome erasure we obtain
\begin{equation}\label{p_0}
    p_0 = 1 - (1-p_{loss}) \cdot \left( 1 - \frac{(1-p_{succ})}{2} \right).
\end{equation}

\begin{figure}[t]
\includegraphics[width=3.3in]{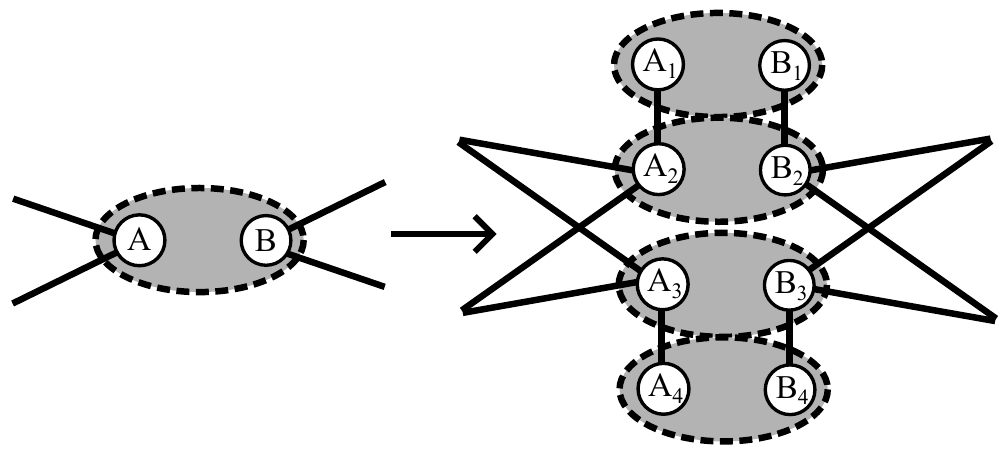}
\caption{\label{fig:shore_encoding} (2,2)-Shor encoding of qubits in a 6-ring state. A single qubit A is encoded by four qubits $\text{A}_i$, same for B. A single fusion measurement is replaced by four, and the result of the encoded measurement is reconstructed from the results of four physical measurements.}
\end{figure}

Redundant resource state encoding enhances error tolerance. For example, each qubit in a graph state may be encoded with a (2,2)-Shor code \cite{Shor1995}. For instance, for a 6-ring state modified by a (2,2)-Shor code as displayed in Fig.~\ref{fig:shore_encoding}, the corresponding measurement erasure probability becomes
\begin{equation}\label{p_enc}
    p_{enc}=\frac{\left( 1-\left(1-p_0\right)^2\right)^2+1-\left(1-p_0^2\right)^2}{2},
\end{equation}
and $p_{enc} > p_0$ if $p_0 > 0.5$.

The loss tolerance of the fusion networks might be further enhanced by the dynamic bias arrangement \cite{https://doi.org/10.48550/arxiv.2303.16122}. This method assumes the choice of a single measurement basis in the case of circuit failure based on the results of previous measurements instead of a random choice.

In the linear optical case, this technique may be improved for some fusion networks because BSM circuits that measure different operators consist of different numbers of basic elements (see Fig.~\ref{fig:table}), and, therefore, possess different losses.

\section{Account of losses}\label{sec:losses}
\subsection{Sources of imperfections in linear optics}

Generally, photons go through three main stages during any linear optical computation: single-photon states are generated, transformed in a linear-optical interferometer, and then detected. Different errors can occur on each stage. The generation quality is characterized by the probability of single-photon generation, purity of single-photon states, and their indistinguishability. Optical losses and inaccuracy of operations describe the transformation stage, and the detection stage error sources include finite detection efficiency, dark counts, detector dead time, jitter and non-ideal photon number resolution capabilities.

Models of optical losses in linear-optical circuits were studied in \cite{Oszmaniec2018, Oh2021,Goldberg2024}, and models of indistinguishability were covered in \cite{Shchesnovich2015, Menssen2017}, for example. Impact of distinguishability errors on notable linear-optical circuits was studied in \cite{Saied2024}. Hardware requirements for realizing quantum advantage through boson sampling for these imperfections along with detection and photon generation inefficiencies in linear-optical schemes were estimated in such works as \cite{https://doi.org/10.48550/arxiv.1809.01953, https://doi.org/10.48550/arxiv.2310.10185}. Fundamental limits for losses in linear optical logical BSMs were studied in works like \cite{Lee2019, https://doi.org/10.48550/arxiv.2302.07908}. There are also investigations of optimal multiphoton entangled state generation circuits in terms of optical losses and indistinguishability, such as \cite{https://doi.org/10.48550/arxiv.2305.08452}.

Here we analyse hardware requirements for linear-optical BSM circuits for fault-tolerant quantum computations in the FBQC model. We consider errors caused by nondeterministic generation of photonic states, optical losses during state transformation, and finite detection efficiency. 

Interferometers composed of linear optical elements -- beamsplitters and phase shifts -- implement the required linear optical transformations. State-of-the-art physical realizations are based on integrated photonic circuits \cite{Elshaari2020, Pelucchi2021, Moody2022, Taballione2019, https://doi.org/10.48550/arxiv.2308.13452}. For example, these can be made on silicon nitride platform \cite{Doerr2006}. Inside the integrated photonic circuits photons may be lost while freely traveling in isolated waveguides (propagation loss), or while passing through nontrivial multimode elements such as beamsplitters, which introduce additional losses.
We will account losses in these two cases separately. All imperfections to be considered may be covered by the beamsplitter loss model.

\subsection{Beamsplitter loss model}
A common way of modeling losses in quantum optics is the beamsplitter loss model \cite{Oszmaniec2018, Oh2021,Goldberg2024}. In this model whenever a photon diverts to a loss channel, it is destroyed with a probability $(1-\eta)$. Thus, for example, after passing through the loss channel, the single-mode optical Fock state $|n\rangle$ consisting of $n$ photons transforms into the state $|r\rangle$ consisting of $r$ photons with probability

\begin{equation}
P(|n\rangle \rightarrow |r\rangle \mid \eta)=\binom{n}{r}\eta^{r}(1-\eta)^{n-r}. 
\end{equation}

\begin{figure}[ht]
    \includegraphics[scale=0.2]{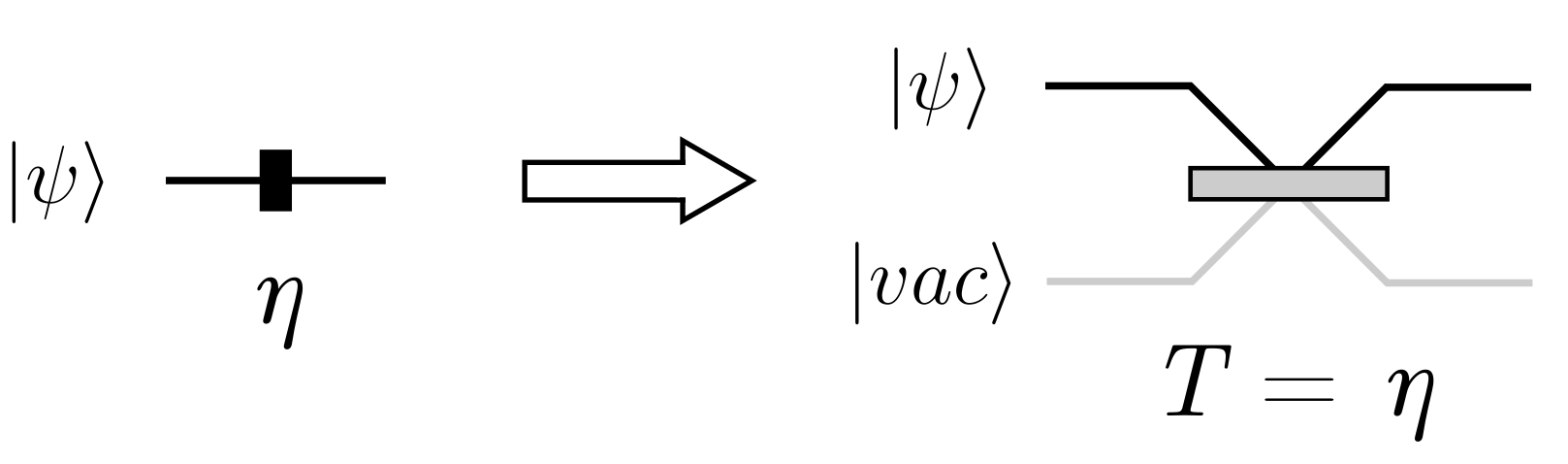}
    \caption{Beam splitter loss model. A loss channel with loss probability $(1-\eta)$ can be replaced by a beamsplitter with transmissivity $\eta$, which reflects the light to some inaccessible (initially not occupied) mode.}
    \label{fig:loss_channel}
\end{figure}

This model is called a beamsplitter loss model because a loss channel can be replaced by a beamsplitter with a transmission probability $\sqrt{\eta}$ and vacuum state at the second input, which reflects the light to some inaccessible mode (see Fig.~\ref{fig:loss_channel}).

\begin{figure}[h]
    \includegraphics[width=3.3in]{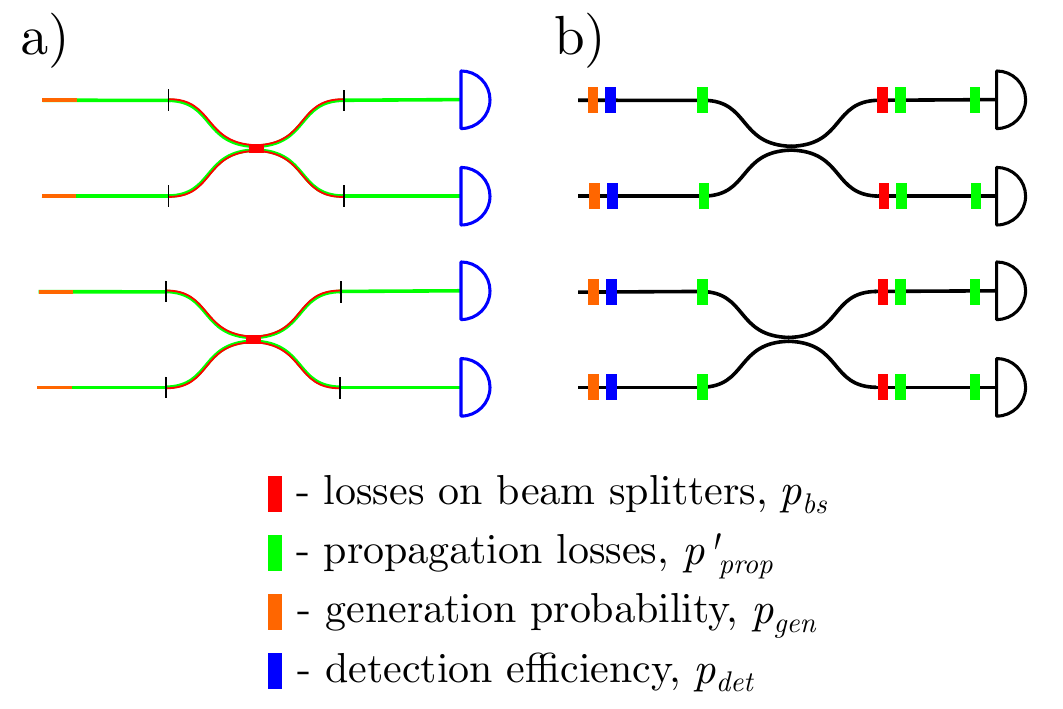}
    \caption{Circuit loss modeling example. Losses from different sources (a) are modeled as corresponding loss channels (b). Loss channels describing detection efficiency may be extracted to the beginning of the scheme.}
    \label{fig:loss_model}
\end{figure}

To examine the action of all sources of loss, we add the corresponding loss channels to the ideal circuits as in the Fig.~\ref{fig:loss_model}. We will assume that the resource state generator emits different photons of a state with equal probability $p_{gen}$, and detection efficiency of each PNRD is $p_{det}$. Using loss channel properties \cite{Oszmaniec2018}, we can shift all loss channels, that describe finite generation probability and detection efficiency, either to the end or to the beginning of the scheme, and combine them together. In this case the new composite single mode loss channel will have the transmission efficiency $\eta = p_{gen} \cdot p_{det} = p_{eff}$, where $p_{eff}$ is a combined generation and detection efficiency.

We consider schemes in which functional elements such as beamsplitters are placed in layers. We suppose that each photon travels approximately the same distance passing through each layer and, thus, the propagation losses $p'_{prop}$ are supposed to be the same for every photon in each layer. To connect our considerations to real hardware we will calculate the limits for specific propagation losses, measured in dB/cm. To estimate these specific values let the length of one layer be $l=500$~\textmu m, which is approximately equal to the corresponding average length in integrated photonic schemes \cite{integrated_photonics}. These assumptions can be easily adopted to a particular practical architecture of an optical circuit, and specific propagation loss values $p_{prop}$ in the following text are connected with corresponding loss probabilities of loss channels modelling propagation losses $p'_{prop}$ as $p'_{prop} = p_{prop} \cdot l$. We also assume that beamsplitters are identical and photons experience the same level of additional losses $p_{bs}$. 

We do not include mode swaps in loss sources, because they are implemented in integrated photonic schemes by waveguide crossings and contribute relatively small loss to the circuits (for example, $43 \pm 4$ mdB per crossing in \cite{Johnson2020}). Moreover, one can eliminate most of the mode swaps in the studied circuits by rearrangement of initial modes and detectors.

To simulate the action of losses we first build a scheme of the lossy circuit from an ideal one as in Fig.~\ref{fig:loss_model}. Then we calculate the unitary transformation matrix of the circuit in an extended space, which includes extra modes for loss channels in the beamsplitter loss model, to account for nonuniform losses. Details are presented in Appendix~\ref{calculations}.

\begin{figure}[t]
    \includegraphics[width=3.33in]{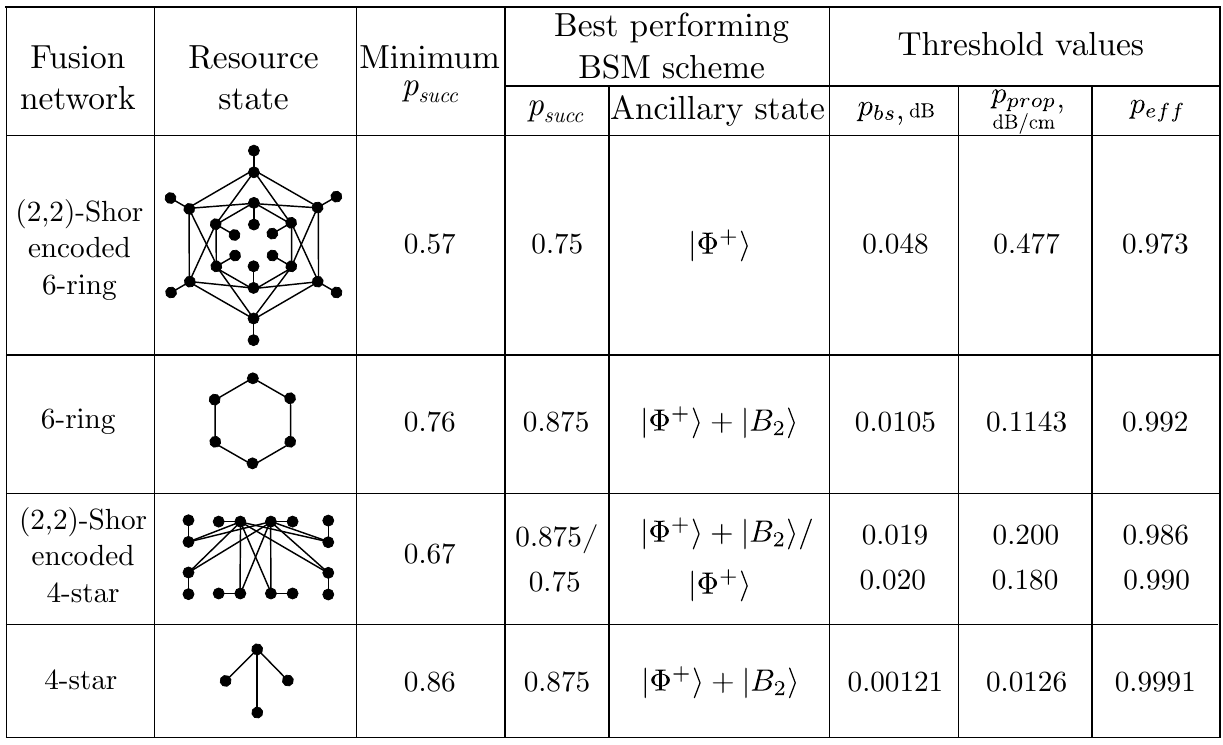}
    \caption{Results of the best-found fusion measurement circuits for studied fusion networks. Resource states are presented in the graph representation up to the single-qubit transformations. Last columns show marginal threshold losses values for correctable regions.  Ancillary states $|\Phi^{+} \rangle = \frac{1}{\sqrt{2}}(|1010\rangle +|0101\rangle)$, $|B_2\rangle = \frac{1}{\sqrt{2}}(|10101010\rangle + |01010101\rangle)$ in the Fock state notation.}
    \label{fig:result_table}
\end{figure}

\begin{figure*}[!ht]
    \includegraphics[width=7in]{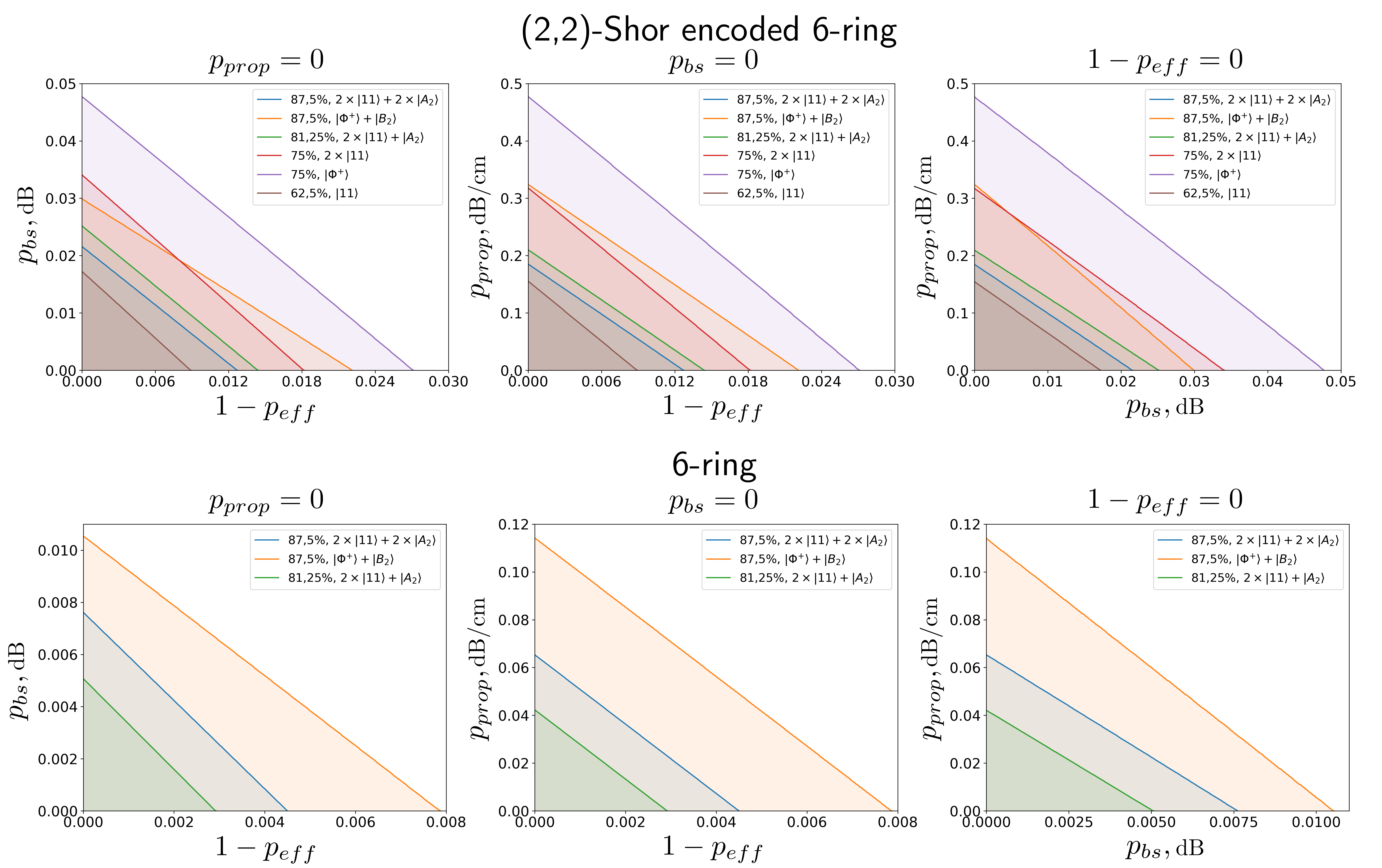}
    \caption{Performance of different BSM schemes used as fusion measurements in the 6-ring and (2,2)-Shor encoded 6-ring fusion networks considering errors caused by nonuniform generation, detection efficiency, propagation losses and losses on beam splitters. Each graph shows slices of 3-dimensional correctable regions with one loss parameter equals to zero. BSM schemes are denoted by their success probability in the lossless case $p_{succ}$ and ancillary states used for it's boosting in the dual-rail encoding, $|\Phi^{+} \rangle = \frac{1}{\sqrt{2}}(|1010\rangle +|0101\rangle)$, $|A_2\rangle = \frac{1}{\sqrt{2}}(|2020\rangle +|0202\rangle)$, $|B_2\rangle = \frac{1}{\sqrt{2}}(|10101010\rangle + |01010101\rangle)$ in the Fock state notation.}
    \label{fig:encoded6star}
\end{figure*}

\section{Lossy BSM}

\begin{figure*}[!ht]
    \includegraphics[width=7in]{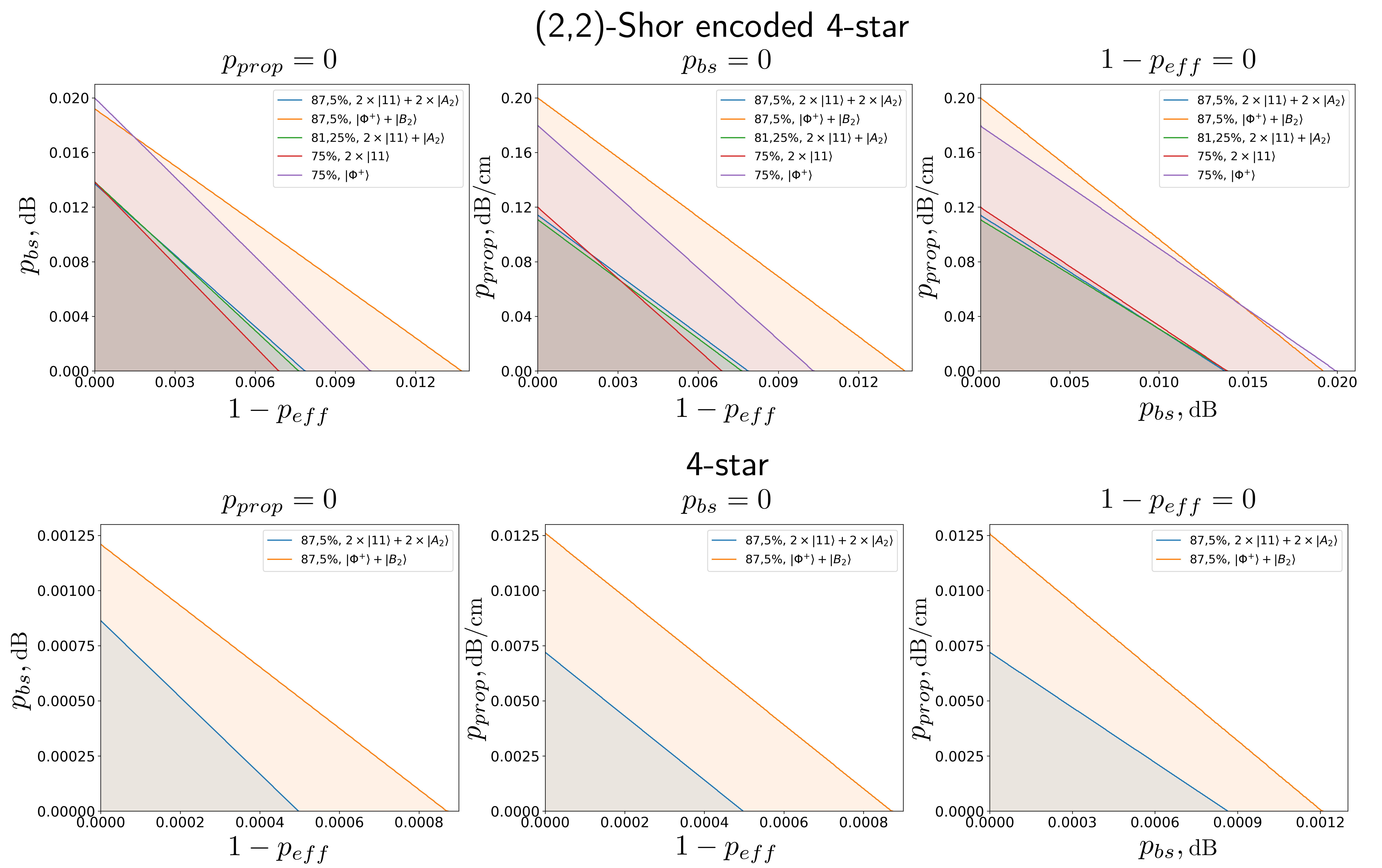}
    \caption{Performance of different BSM schemes used as fusion measurements in the 4-star and (2,2)-Shor encoded 4-star fusion networks considering errors caused by nonuniform generation, detection efficiency, propagation losses and losses on beam splitters. Each graph shows slices of 3-dimensional correctable regions with one loss parameter equals to zero. BSM schemes are denoted by their success probability in the lossless case $p_{succ}$ and ancillary states used for it's boosting in the dual-rail encoding, $|\Phi^{+} \rangle = \frac{1}{\sqrt{2}}(|1010\rangle +|0101\rangle)$, $|A_2\rangle = \frac{1}{\sqrt{2}}(|2020\rangle +|0202\rangle)$, $|B_2\rangle = \frac{1}{\sqrt{2}}(|10101010\rangle + |01010101\rangle)$ in the Fock state notation.}
    \label{fig:full4star}
\end{figure*}

We consider the performance of linear-optical BSM circuits within fusion networks based on 6-ring and 4-star together with an optional (2,2)-Shor encoding. As it has been noted in Section~\ref{sec:FBQC}, errors in fusion networks may be corrected if the measurement erasure probability $p_0$ ($p_{enc}$ if states are (2,2)-Shor encoded) is below the threshold value for the chosen fusion network. The $p_0$ depends on the BSM success probability $p_{succ}$ in the lossless case and the probability that at least one photon has been lost during the action of the circuit $p_{loss}$:
\begin{equation}
    p_0 = p_0(p_{succ}, p_{loss}).
\end{equation}

This equation causes a trade-off: $p_{succ}$ increases with the number of ancillary photons used in the boosted BSM circuit, which leads to decrease of erasure probability. However, $p_{loss}$ also increases with the number of ancillary photons, as well as with the depth of the circuit and the number of active elements, suppressing the improvement obtained from higher $p_{succ}$.

We compare circuits with different $p_{succ}$ and different resource states used for boosting. Results for the best performing schemes are presented in Fig.~\ref{fig:result_table}, where we show marginal threshold values for losses from different sources below which the corresponding errors are correctable within the fusion networks. 

As is was mentioned before, error tolerance benefits from the dynamic bias arrangement, but here we give a lower bound on the BSM circuit performance and consider only static bias arrangement. Besides, although generally $p_{loss}$ differs with the number of basic elements in the circuit, we assume that the circuits that measure $XX$ or $ZZ$ operators in the case of failure have the same $p_{loss}$, the highest one for both cases (see. Fig~\ref{fig:table}).

\subsection{6-ring fusion network}

In \cite{Bartolucci2023} the authors demonstrated that in the marginal case when the probability of every measurement result flip in a 6-ring fusion network equals to zero, errors are in the correctable region of the fusion network if the measurement erasure probability lies below the threshold value of $p_{er} = 0.12$. Hence, the loss threshold stems from the fact that the probability of any photon loss $p_{loss}$ for a given BSM circuit architecture together with its success probability in the lossless case $p_{succ}$ should yield the $p_0$ value below the threshold $p_{er}$ (see formulas (\ref{p_0}) and (\ref{p_enc})).

To stay below threshold for the 6-ring fusion network, one needs to use BSM circuits with $p_{succ}$ higher than $0.76$, or $0.57$ in the (2,2)-Shor encoded network even in the absence of losses according to formulas (\ref{p_0}) and (\ref{p_enc}).

We consider BSM circuits of the form shown in Fig.~\ref{bsm_table}. Fig.~\ref{fig:encoded6star} indicates correctable areas limited by threshold surfaces for different BSM circuits used as fusion measurements in 6-ring-based fusion networks. The fusion network can tolerate errors induced by losses in a certain BSM circuit if a combination of loss parameters defines the coordinates of a point on the graph that lies inside the correctable region limited by the threshold surface for the corresponding circuit. Circuits in Fig.~\ref{fig:encoded6star} are denoted by their $p_{succ}$ probability and ancillary states used for the boosting in the Fock state notation.

It can be seen that for the (2,2)-Shor encoded network the scheme with $p_{succ} = 0.75$ and $|\Phi^{+} \rangle = \frac{1}{\sqrt{2}}(|1010\rangle +|0101\rangle)$ as the ancillary state shows the best results, and further success probability increase does not enhance the performance. Marginal loss thresholds for this scheme are shown in Fig.~\ref{fig:result_table}.
However, schemes with $|11\rangle$ and $2\times |11\rangle$ ancillary states might also be useful since their ancillary states are not entangled and, hence, easier to create.

For the 6-ring fusion network the best results are demonstrated by the scheme with $p_{succ} = 0.875$ and $|\Phi^{+} \rangle$ and $|B_2\rangle = \frac{1}{\sqrt{2}}(|10101010\rangle + |01010101\rangle)$ as ancillary states. The marginal loss thresholds for this scheme are presented in Fig.~\ref{fig:result_table}. Further success probability boosting seems unprofitable, because of the escalating complexity of ancillary states.

\subsection{4-star fusion network}

Minimal success probabilities $p_{succ}$ that yield the $p_0$ value less than the threshold value of $p_{er} = 0.069$ \cite{Bartolucci2023} for the 4-star-based fusion networks are 0.86 for the 4-star fusion network and 0.67 for the (2,2)-Shor encoded network.

Fig.~\ref{fig:full4star} reveals that for a (2,2)-Shor encoded network schemes with 0.875 and 0.75 success probabilities and corresponding ancillary states $|\Phi^{+} \rangle + |B_2\rangle$ and $|\Phi^{+} \rangle$ show the best results for different relations of losses values. The later circuit slightly outperforms the former in the high $p_{bs}$ scenario. Marginal losses thresholds for these schemes are presented in Fig.~\ref{fig:result_table}.

For the 4-star fusion network it is the circuit with 0.875 success probability and $|\Phi^{+} \rangle + |B_2\rangle$ ancillary states which demonstrates the best results. However,
hardware requirements for this scheme are much higher than for the others, see Fig. \ref{fig:result_table}.

\section{Discussion}

Our simulation evidences that if we are able to efficiently create resource states for fusion networks and ancillary states for boosted BSM schemes and possess PNRDs with cumulative detection efficiency higher than approximately 0.97, the photonic linear-optical BSM circuits with losses less than 0.048 dB per individual beam splitters and 0.48 dB/cm for propagation enable error-tolerant quantum computing using (2,2)-Shor encoded 6-ring fusion network in the FBQC model. Modern integrated photonic technology comes close to these limits. For example the $8\times8$ reconfigurable interferometer fabricated using the SiN platform \cite{Taballione2019} exhibits 0.2 dB/cm propagation losses.

With state-of-the-art single-photon detectors reaching the detection efficiency of 0.98 \cite{Reddy2020} the realization of fusion measurements within the linear-optical implementation of the FBQC model is feasible with the current level of technologies.

However, both single-photon generation \cite{https://doi.org/10.48550/arxiv.2306.00874, https://doi.org/10.48550/arxiv.2307.02189, https://doi.org/10.48550/arxiv.2302.06282} and resource state generation \cite{https://doi.org/10.48550/arxiv.2106.13825, Gubarev2020, Stanisic2017, Fldzhyan2021, Fldzhyan2023} remain a big problem in linear optics. According to our estimates, for example, for 0.98 detection efficiency the probability of generation of a single photon of a resource state should be higher than approximately 0.99 for the (2,2)-Shor encoded 6-ring fusion network, which seems far from being achievable. Nevertheless, it must be noted that optimal ways of generating linear-optical resource states, best-performing fusion networks and circuits for BSMs have not yet been found.

The BSM fusion measurement performance in FBQC may be further improved by developing other approaches to boosting of the BSM schemes, using other fusion networks or more efficient state encoding \cite{Bell2023} and application of the dynamic bias arrangement \cite{https://doi.org/10.48550/arxiv.2303.16122}.

Our results allow to estimate the prospects of linear-optical realization of quantum computations in the FBQC model.
Circuit architectures and their simulation results presented here may shed light onto new aspects of linear-optical circuit design and fuel further search for an optimized BSM circuit.

\section{Acknowledgements}

We would like to thank Mikhail Saygin for fruitful discussions and careful reading of the manuscript.

The work of authors was supported by Rosatom in the framework of the Roadmap for Quantum computing (Contract No. 868-1.3-15/15-2021 dated October 5, 2021 and Contract No.P2124 dated November 24,2021)

I.D. acknowledges support from  Russian Science Foundation grant 22-12-00353 (https://rscf.ru/en/project/22-12-00353/).

\bibliographystyle{unsrt}
\bibliography{biblio}

\appendix

\section{Numerical simulation}\label{calculations}

Using the assumptions mentioned in Section~\ref{sec:losses}, we construct lossy BSM circuits as in Fig.~\ref{fig:loss_model}. Then we calculate the transformation of the initial state imposed by the circuit. In the ideal lossless case if the input to the scheme is an $M$-mode $N$-photon Fock state
\begin{equation}
		\label{in}
		|\psi_{\mathrm{in}} \rangle=| n_{1}, n_{2}, \ldots, n_{M}\rangle=\prod_{k=1}^{M} \frac{\left(\hat{a}_{k}^{\dagger}\right)^{n_{k}}}{\sqrt{n_{k} !}} | \mathrm{vac} \rangle,
\end{equation}
where $| \mathrm{vac} \rangle=| 0 \rangle^{\otimes M}$, the transformation is described by 
\begin{align}
		\mathcal{U} | \psi_{\mathrm{in}} \rangle&=\prod_{k=1}^{M} \frac{1}{\sqrt{n_{k} !}} (\mathcal{U} \hat{a}_{k}^{\dagger} \mathcal{U}^{\dagger} )^{n_{k}} \mathcal{U} | \mathrm{vac} \rangle \\
		&=\prod_{k=1}^{M} \frac{1}{\sqrt{n_{k} !}}\left(\sum_{j=1}^{M} \hat{a}_{j}^{\dagger} U_{j k}\right)^{n_{k}} | \mathrm{vac} \rangle,
\end{align}
where $U_{j k}$ are matrix elements of the corresponding optical circuit matrix $U$. Then the coefficients within the final state are determined by the formula 
\begin{equation}\label{out}
		\langle\phi|\mathcal{U}| \psi\rangle=\frac{\operatorname{perm}\left(U_{\psi \phi}\right)}{\sqrt{n_{1} ! \cdots n_{M} ! m_{1} ! \cdots m_{M} !}},
\end{equation}
where $U_{\psi \phi}$ is the matrix assembled from the elements of the matrix $U$ (see \cite{https://doi.org/10.48550/arxiv.quant-ph/0406127} for a detailed explanation).

We account for the losses using a following technique. If we know how linear optical elements and loss channels are located in the circuit, we can at first calculate the number of loss channels $M_{loss}$ and consider the initial state in the extended space. If the initial state is an $M$-mode Fock state, in the extended space it will be 
\begin{equation} \label{subsystems}
		| \widetilde{\psi}_{in} \rangle =| n_{1}, n_{2}, \ldots, n_{M}, 0_{M+1}, \ldots, 0_{M+M_{loss}}\rangle,
\end{equation}
where $(M+1), \ldots, (M+M_{loss})$ are extra loss modes. Then we can calculate the unitary transformation matrix in the extended space $ \widetilde{U}$ by multiplying matrices of each element according to their placement in the circuit. Loss channels are modeled as beam splitters that reflect the light to extra loss modes here. Having obtained the output state in the extended space, we can take a partial trace over extra loss modes to calculate the output of the lossy circuit in the original space. The output state thus will be a density matrix, and we may derive the probability of at least one photon loss $p_{loss}$ as $1-$ the coefficient of $N$-photon constituent of the density matrix.

As long as we are only interested in the probability of all photons' survival $p_{surv}$, which is equivalent to finding the probability of at least one photon loss, we can instead construct a non-unitary matrix $L$ describing a certain lossy circuit in the original space by considering loss channels as non-unitary elements that reduce amplitudes of the Fock states passing through them. Then, we can take the state of two initially unentangled photonic dual-rail encoded qubits along with the corresponding ancillary state as an input to the lossy BSM circuit, quantify output amplitudes by the formula (\ref{out}), replacing $U$ with $L$ and derive the probability of all photons' survival  $p_{surv}$. The probability of at least one photon loss will be then $p_{loss} = 1-p_{surv}$. This second technique will not give us full output states of lossy circuits in the density matrix form, but it significantly reduces the calculation time.

Obtaining $p_{loss}$, we determine the measurement erasure probability $p_{0}$ ($p_{enc}$ in the encoded case) for the specific circuit as $p_0 = p_0(p_{succ}, p_{loss})$ according to formulas (\ref{p_0}) and (\ref{p_enc}).

\end{document}